\documentclass{nature}
\usepackage{graphicx}
\usepackage{amsmath}
\usepackage{amssymb}
\usepackage{color}
\usepackage{url}
\usepackage{verbatim}
\usepackage{tabu}
\usepackage[normalem]{ulem}
\usepackage{sansmath}
\usepackage{upgreek}
\usepackage{epsfig}
\usepackage[left]{lineno}
\usepackage{setspace}
\usepackage{caption}
\usepackage{url}
\usepackage[dvipsnames]{xcolor}

\usepackage[margin=1.1in]{geometry}

\makeatletter
\let\blx@rerun@biber\relax
\makeatother

\DeclareCaptionLabelFormat{adja-page}{\hrulefill\\#1 #2 \emph{(previous page)}}

\DeclareSourcemap{
  \maps[datatype=bibtex]{
    \map[overwrite]{
      \step[fieldsource=month, match=\regexp{\A(j|J)an(uary)?\Z}, replace=1]
      \step[fieldsource=month, match=\regexp{\A(f|F)eb(ruary)?\Z}, replace=2]
      \step[fieldsource=month, match=\regexp{\A(m|M)ar(ch)?\Z}, replace=3]
      \step[fieldsource=month, match=\regexp{\A(a|A)pr(il)?\Z}, replace=4]
      \step[fieldsource=month, match=\regexp{\A(m|M)ay\Z}, replace=5]
      \step[fieldsource=month, match=\regexp{\A(j|J)un(e)?\Z}, replace=6]
      \step[fieldsource=month, match=\regexp{\A(j|J)ul(y)?\Z}, replace=7]
      \step[fieldsource=month, match=\regexp{\A(a|A)ug(ust)?\Z}, replace=8]
      \step[fieldsource=month, match=\regexp{\A(s|S)ep(tember)?\Z}, replace=9]
      \step[fieldsource=month, match=\regexp{\A(o|O)ct(ober)?\Z}, replace=10]
      \step[fieldsource=month, match=\regexp{\A(n|N)ov(ember)?\Z}, replace=11]
      \step[fieldsource=month, match=\regexp{\A(d|D)ec(ember)?\Z}, replace=12]
    }
  }
}

\def\lapp{\ifmmode\stackrel{<}{_{\sim}}\else$\stackrel{<}{_{\sim}}$\fi}
\def\gapp{\ifmmode\stackrel{>}{_{\sim}}\else$\stackrel{>}{_{\sim}}$\fi}

\defbibheading{subbibliography}{%
  \section*{}}

\let\cite\autocite

\title{Sub-second periodicity in a fast radio burst}

\author{The CHIME/FRB Collaboration:
Bridget C.~Andersen$^{1,2}$, \allowbreak
Kevin Bandura$^{3,4}$, \allowbreak
Mohit Bhardwaj$^{1,2}$, \allowbreak
P.~J.~Boyle$^{1,2}$, \allowbreak
Charanjot Brar$^{1,2}$, \allowbreak
Daniela Breitman$^{5,6,7}$, \allowbreak
Tomas Cassanelli$^{6,7}$, \allowbreak
Shami Chatterjee$^{8}$, \allowbreak
Pragya Chawla$^{1,2}$, \allowbreak
Jean-François Cliche$^{1,2}$, \allowbreak
Davor Cubranic$^{9}$, \allowbreak
Alice P.~Curtin$^{1,2}$, \allowbreak
Meiling Deng$^{10,11}$, \allowbreak
Matt Dobbs$^{1,2}$, \allowbreak
Fengqiu Adam Dong$^{9}$, \allowbreak
Emmanuel Fonseca$^{1,2}$, \allowbreak
B.~M.~Gaensler$^{6,7}$, \allowbreak
Utkarsh Giri$^{10,12}$, \allowbreak
Deborah C.~Good$^{9}$, \allowbreak
Alex S.~Hill$^{13,11}$, \allowbreak
Alexander Josephy$^{1,2}$, \allowbreak
J.~F.~Kaczmarek$^{11}$, \allowbreak
Zarif Kader$^{1,2}$, \allowbreak
Joseph Kania$^{14,4}$, \allowbreak
Victoria M.~Kaspi$^{1,2}$, \allowbreak
Calvin Leung$^{15,16}$, \allowbreak
D.~Z.~Li$^{17}$, \allowbreak
Hsiu-Hsien Lin$^{18,19}$, \allowbreak
Kiyoshi W.~Masui$^{15,16}$, \allowbreak
Ryan Mckinven$^{7,6}$, \allowbreak
Juan Mena-Parra$^{15}$, \allowbreak
Marcus Merryfield$^{1,2}$, \allowbreak
B.~W.~Meyers$^{9}$, \allowbreak
D.~Michilli$^{1,2,15,16*}$, \allowbreak
Arun Naidu$^{1,2}$, \allowbreak
Laura Newburgh$^{20}$, \allowbreak
C.~Ng$^{6}$, \allowbreak
Anna Ordog$^{13,11}$, \allowbreak
Chitrang Patel$^{1,6}$, \allowbreak
Aaron B.~Pearlman$^{1,2}$, \allowbreak
Ue-Li Pen$^{18,19,21,6,10}$, \allowbreak
Emily Petroff$^{1,2,22}$, \allowbreak
Ziggy Pleunis$^{1,2}$, \allowbreak
Masoud Rafiei-Ravandi$^{10,12}$, \allowbreak
Mubdi Rahman$^{23}$, \allowbreak
Scott Ransom$^{24}$, \allowbreak
Andre Renard$^{6}$, \allowbreak
Pranav Sanghavi$^{3,4}$, \allowbreak
Paul Scholz$^{6}$, \allowbreak
J.~Richard Shaw$^{9}$, \allowbreak
Kaitlyn Shin$^{15,16}$, \allowbreak
Seth R.~Siegel$^{1,2}$, \allowbreak
Saurabh Singh$^{1,2}$, \allowbreak
Kendrick Smith$^{10}$, \allowbreak
Ingrid Stairs$^{9}$, \allowbreak
Chia Min Tan$^{1,2}$, \allowbreak
Shriharsh P.~Tendulkar$^{25,26}$, \allowbreak
Keith Vanderlinde$^{6,7}$, \allowbreak
D.~V.~Wiebe$^{9}$, \allowbreak
Dallas Wulf$^{1,2}$, \allowbreak
Andrew Zwaniga$^{1,2}$ \allowbreak
}
\newcommand{\affils}{
\begin{affiliations}
\item{Department of Physics, McGill University, 3600 rue University, Montr\'eal, QC H3A 2T8, Canada}
\item{McGill Space Institute, McGill University, 3550 rue University, Montr\'eal, QC H3A 2A7, Canada}
\item{Lane Department of Computer Science and Electrical Engineering, 1220 Evansdale Drive, PO Box 6109, Morgantown, WV 26506, USA}
\item{Center for Gravitational Waves and Cosmology, West Virginia University, Chestnut Ridge Research Building, Morgantown, WV 26505, USA}
\item{Department of Physics, University of Toronto, 60 St.~George Street, Toronto, ON M5S 1A7, Canada}
\item{Dunlap Institute for Astronomy \& Astrophysics, University of Toronto, 50 St.~George Street, Toronto, ON M5S 3H4, Canada}
\item{David A.~Dunlap Department of Astronomy \& Astrophysics, University of Toronto, 50 St.~George Street, Toronto, ON M5S 3H4, Canada}
\item{Cornell Center for Astrophysics and Planetary Science, Cornell University, Ithaca, NY 14853, USA}
\item{Department of Physics and Astronomy, University of British Columbia, 6224 Agricultural Road, Vancouver, BC V6T 1Z1 Canada}
\item{Perimeter Institute for Theoretical Physics, 31 Caroline Street N, Waterloo, ON N25 2YL, Canada}
\item{Dominion Radio Astrophysical Observatory, Herzberg Research Centre for Astronomy and Astrophysics, National Research Council Canada, PO Box 248, Penticton, BC V2A 6J9, Canada}
\item{Department of Physics and Astronomy, University of Waterloo, Waterloo, ON N2L 3G1, Canada}
\item{Department of Computer Science, Math, Physics, \& Statistics, University of British Columbia, Okanagan Campus, Kelowna, BC V1V 1V7, Canada}
\item{Department of Physics and Astronomy, West Virginia University, PO Box 6315, Morgantown, WV 26506, USA }
\item{MIT Kavli Institute for Astrophysics and Space Research, Massachusetts Institute of Technology, 77 Massachusetts Ave, Cambridge, MA 02139, USA}
\item{Department of Physics, Massachusetts Institute of Technology, 77 Massachusetts Ave, Cambridge, MA 02139, USA}
\item{Cahill Center for Astronomy and Astrophysics, California Institute of Technology, 1216 E California Boulevard, Pasadena, CA 91125, USA}
\item{Institute of Astronomy and Astrophysics, Academia Sinica, Astronomy-Mathematics Building, No. 1, Sec. 4, Roosevelt Road, Taipei 10617, Taiwan}
\item{Canadian Institute for Theoretical Astrophysics, 60 St.~George Street, Toronto, ON M5S 3H8, Canada}
\item{Department of Physics, Yale University, New Haven, CT 06520, USA}
\item{Canadian Institute for Advanced Research, MaRS Centre, West Tower, 661 University Ave, Suite 505, Toronto, ON, M5G 1M1 Canada}
\item{Anton Pannekoek Institute for Astronomy, University of Amsterdam, Science Park 904, 1098 XH Amsterdam, The Netherlands}
\item{Sidrat Research, PO Box 73527 RPO Wychwood, Toronto, ON M6C 4A7, Canada}
\item{National Radio Astronomy Observatory, 520 Edgemont Rd, Charlottesville, VA 22903, USA}
\item{Department of Astronomy and Astrophysics, Tata Institute of Fundamental Research, Mumbai, 400005, India}
\item{National Centre for Radio Astrophysics, Post Bag 3, Ganeshkhind, Pune, 411007, India}
\end{affiliations}
}
\newcommand{\allacks}{
A.B.P. is a McGill Space Institute (MSI) Fellow and a Fonds de Recherche du Quebec -- Nature et Technologies (FRQNT) postdoctoral fellow.
A.O. is supported by the Dunlap Institute.
A.S.H. is supported by an NSERC Discovery Grant.
B.M.G. is supported by an NSERC Discovery Grant (RGPIN-2015-05948), and by the Canada Research Chairs (CRC) program. 
C.L. was supported by the U.S. Department of Defense (DoD) through the National Defense Science \& Engineering Graduate Fellowship (NDSEG) Program.
D.C.G. is supported by the John I. Watters Research Fellowship.
D.M. is a Banting Fellow.
E.P. acknowledges funding from an NWO Veni Fellowship.
J.M.P is a Kavli Fellow.
K.B. is supported by an NSF grant (2006548).
K.W.M. is supported by an NSF Grant (2008031).
M.B. is supported by an FRQNT Doctoral Research Award.
M.D. is supported by a Killam Fellowship, Canada Research Chair, NSERC Discovery Grant, CIFAR, and by the FRQNT Centre de Recherche en Astrophysique du Qu\'ebec (CRAQ).
M.M. is supported by an NSERC PGS-D award.
P.C. is supported by an FRQNT Doctoral Research Award.
P.S. is a Dunlap Fellow and an NSERC Postdoctoral Fellow. 
S.C. acknowledges support from the National Science Foundation (AAG 1815242).
S.M.R. is a CIFAR Fellow and is supported by the NSF Physics Frontiers Center award 1430284.
U.L.P. receives the support of the Natural Sciences and Engineering Research Council of Canada (NSERC), [funding reference number RGPIN-2019-067, CRD 523638-18, 555585-20], Ontario Research Fund—research Excellence Program (ORF-RE), Canadian Institute for Advanced Research (CIFAR), Thoth Technology Inc, Alexander von Humboldt Foundation, and the Ministry of Science and Technology(MOST) of Taiwan(110-2112-M-001-071-MY3). 
V.M.K. holds the Lorne Trottier Chair in Astrophysics \& Cosmology and a Distinguished James McGill Professorship and receives support from an NSERC Discovery Grant and Herzberg Award, from an R. Howard Webster Foundation Fellowship from the Canadian Institute for Advanced Research (CIFAR), and from the FRQNT Centre de Recherche en Astrophysique du Quebec.
}

\begin{document}

\maketitle

\clearpage
\blfootnote{
\hspace{-0.7cm}
* Corresponding author: Daniele Michilli. E-mail: danielemichilli@gmail.com 
\affils
}

\clearpage
\begin{abstract}
Fast radio bursts (FRBs) are millisecond-duration flashes of radio waves that are visible at distances of billions of light-years.\cite{lbm+07}
The nature of their progenitors and their emission mechanism remain open astrophysical questions.\cite{pww+19}
Here we report the detection of the multi-component FRB 20191221A and the identification of a periodic separation of $216.8(1)$\,ms between its components with a significance of $6.5\sigma$.
The long ($\sim 3$\,s) duration and nine or more components forming the pulse profile make this source an outlier in the FRB population. 
Such short periodicity provides strong evidence for a neutron-star origin of the event.
Moreover, our detection favours emission arising from the neutron-star magnetosphere,\cite{pp10,lkz20} as opposed to emission regions located further away from the star, as predicted by some models.\cite{mms19}
\end{abstract}

Operating on the Canadian Hydrogen Intensity Mapping Experiment (CHIME), CHIME/FRB\cite{abb+18} is an ongoing experiment to find and study a large number\cite{1stcatalog} of FRBs.  
CHIME is a cylindrical North-South oriented transit radio interferometer observing in the 400--800-MHz range.
Upon detection of an FRB, the so-called \emph{intensity} data, i.e. the total intensity of the signal as a function of time and frequency, are stored. 
Additionally, channelized complex voltages (referred to as \emph{baseband} data) with full polarisation information are stored for a subset of FRBs (see Methods).

\begin{table}[t!]
\begin{center}
\begin{tabular}{ll}
\hline
Parameter & 20191221A \\
\hline
MJD$^*$ & 58838.20638077, \\
 & 58838.20630684 \\
RA$^*$ J2000 (deg) & 44.6(1), 38.4(2) \\
Dec$^*$ J2000 (deg) & 79.74(2), 79.73(3) \\
$l^*$ (deg) & 128.60(2), 127.56(4) \\
$b^*$ (deg) & 18.30(2), 17.78(3) \\
DM (pc\,cm$^{-3}$) & 368(6) \\
Period (ms) & 216.8(1) \\
Period significance ($\sigma$) & 6.5 \\
Average width (ms) & 4(1) \\
Scattering (ms) & 340(10) \\
Fluence$^{\dagger}$ (Jy ms) & 1.2(4)$\,\times\,$10$^3$ \\
Peak Flux$^{\dagger}$ (Jy) & 2.0(1.0) \\
Exposure$^\ddagger$ (hours) & 340.1(2), 106(4) \\
\hline
\end{tabular}
\end{center}
$^*$ Two localization regions are equally probable and both positions are reported. \\
$^\dagger$ These are lower limits as detailed in the Methods. \\
$^\ddagger$ For circumpolar sources ($\delta > +70^\circ$), the two entries correspond to exposure in the upper and lower transit, respectively. \\
\caption{
{\bf Properties of FRB 20191221A.} 
  Uncertainties are reported at 1-$\sigma$ confidence level.  The arrival time is that of the brightest sub-burst at the Solar System's barycentre and infinite frequency. The DM is calculated to maximize the peak S/N in the timeseries. The scattering timescale is referenced to the centre of the band, i.e. $\sim 600$\,MHz.
  Fluence is for the full band-averaged profile, and peak flux is the maximum in the profile (with 1-ms time resolution). 
\label{tab:bursts}
}
\end{table}

\begin{figure}[htbp]
\begin{center}
\includegraphics[width=0.5\textwidth]{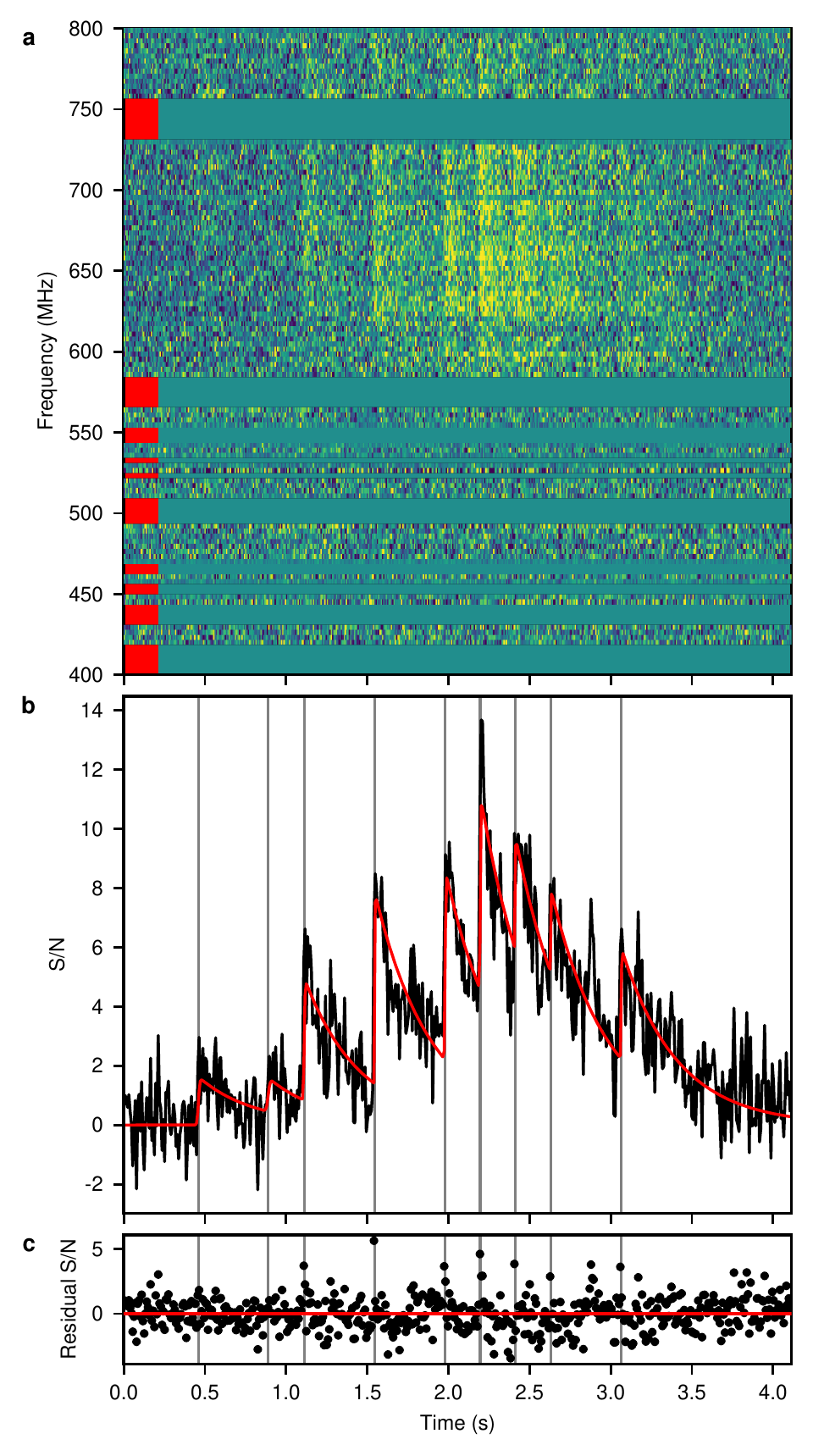}
\caption{
\textbf{Radio signal from FRB 20191221A.}
\textbf{a}, Waterfall plot of the signal intensity (colour-coded) as a function of time and frequency. Frequency channels missing or masked due to radio frequency interference are replaced with off-burst median values and are indicated in red. 
Effects of dispersion have been removed, and data have been averaged to 3.125\,MHz frequency resolution and to 7.86432\,ms time resolution.
\textbf{b}, In black, the pulse profile obtained by averaging the frequency channels of the waterfall plot where signal is visible. The Gaussian function convolved with an exponential used to model nine peaks in the profile is plotted in red, and peak locations are highlighted by vertical lines.
\textbf{c}, Residuals of the fit, with a red horizontal line at zero residual.
}
\label{fig:waterfall}
\end{center}
\end{figure}

In fewer than 0.5\% of the events detected by CHIME/FRB, five or more separate components are visible in the pulse profiles\cite{1stcatalog} obtained by summing all frequency channels of an intensity dataset after correcting for the effects of the dispersion measure (DM).
Particularly striking is FRB 20191221A, with a total duration of roughly three seconds and at least nine overlapping components (Figure~\ref{fig:waterfall} and Table~\ref{tab:bursts}).
No other FRB candidate observed by CHIME/FRB contains a comparable or greater number of sub-components.
Since the detection pipeline of CHIME/FRB is not optimized to find bursts longer than 128\,ms, it is possible that some events with comparable duration eluded detection. However, FRB 20191221A was identified using the individual peaks in its profile. Therefore,  we expect the fraction of missed long-duration events to be small if single peaks can be identified in their pulse profiles.
A detailed analysis of the completeness of the CHIME/FRB search pipeline has been presented elsewhere.\cite{1stcatalog}

\begin{figure}[htbp]
\begin{center}
\includegraphics[width=0.5\textwidth]{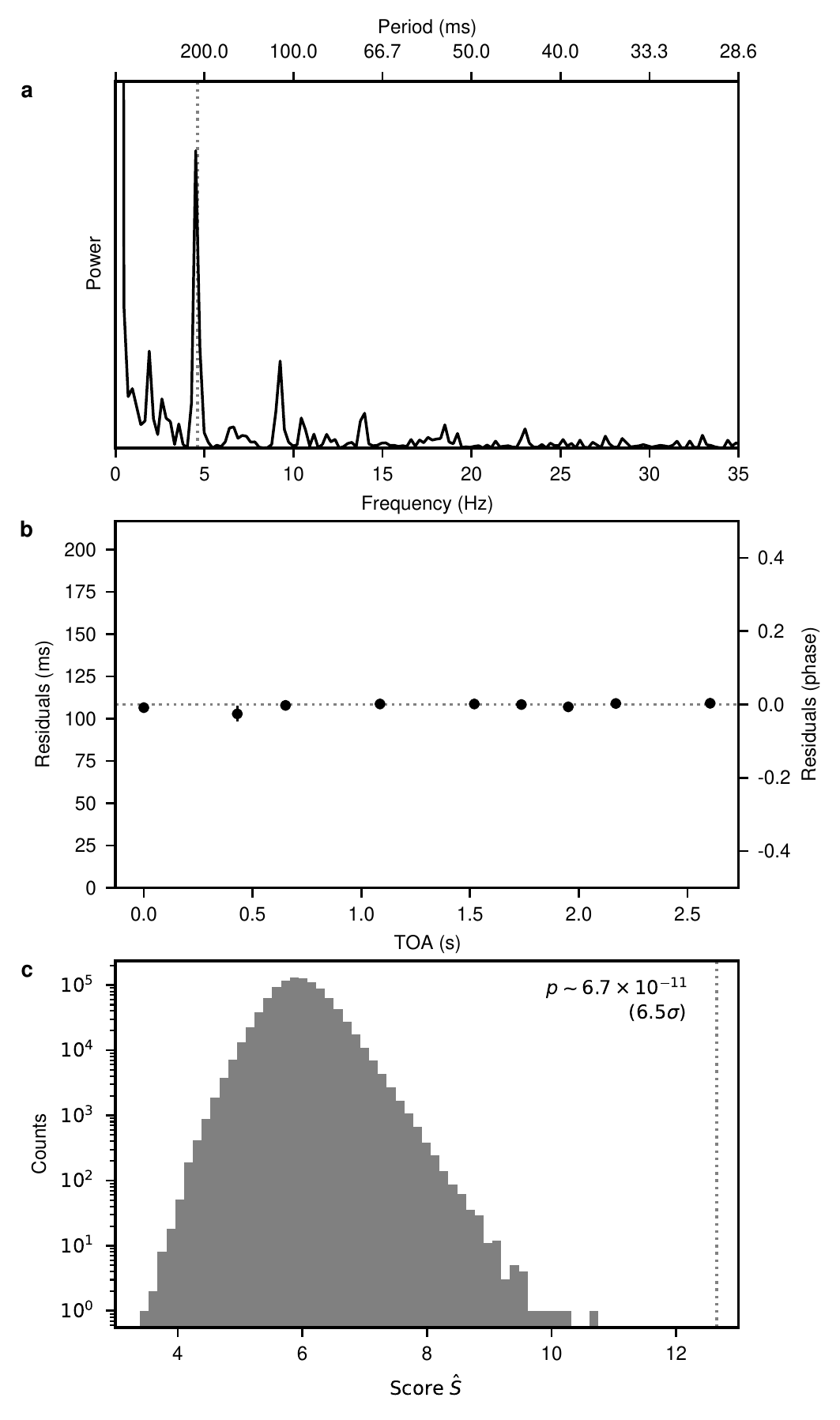}
\caption{
\textbf{Periodicity analysis of FRB 20191221A.}
\textbf{a}, Power spectrum obtained with a discrete Fourier Transform of the pulse profile. The vertical dotted line indicates a period $P = 216.8$\,ms.
\textbf{b}, Residuals of a timing analysis assuming that the peaks forming the FRB profile are separated by integer numbers times a period $P = 216.8$\,ms. 1-$\sigma$ error bars are often smaller than the symbol size. The horizontal dotted line indicates a phase of zero around which residuals have been rotated.
\textbf{c}, Study of the statistical significance of the measured periodicity by using a periodicity-sensitive score $\hat S$ described in the Methods. The grey histogram has been obtained with an ensemble of simulations, whereas the value measured for the FRB is represented with a vertical dotted line. The corresponding probability of obtaining such a periodicity by chance is indicated on the plot.
}
\label{fig:periodicity}
\end{center}
\end{figure}

Significant peaks are visible in the power spectrum of FRB 20191221A obtained by performing a Fast Fourier Transform (FFT) on its pulse profile (Figure~\ref{fig:periodicity}), indicating a possible periodicity in the times of arrival (ToAs) of single components of the pulse profile.
To confirm this, individual sub-components have been fitted with a Gaussian function convolved with a single exponential to account for scattering, a pulse broadening caused by the propagation of the radio waves in turbulent plasma.\cite{ric90}
The resulting ToAs have been used to perform a timing analysis around the initial period derived from the power spectra.
A simple timing model with two free parameters, a period and an arbitrary overall phase, have been fitted to the data. 
The fit residuals are shown in Figure~\ref{fig:periodicity} and the period is reported in Table~\ref{tab:bursts}.
A comparison of the residuals with a straight line, expected for a perfectly periodic source, yields $\chi^2_\textrm{red}=1.09$, indicating that the signal does not deviate significantly from a strict periodicity.
We calculated the significance of the periodicity both by simulating a distribution of random ToAs and by using a Rayleigh periodicity analysis, as described in the Methods.
The spurious detection probability calculated for the $216.8(1)$-ms period of FRB 20191221A is $6.7 \times 10^{-11}$.
Therefore, the periodicity is robust with high confidence.

We found two other notable sources, FRBs 20210206A and 20210213A, that are formed by five and six components and show possible periodicities of 2.8(1) and 10.7(1)\,ms, respectively.
However, the spurious detection probabilities of $0.2$ and $0.02$ measured for their periodicities do not allow us to exclude a chance coincidence. 
Therefore, we do not use them in our analysis but we present them in the Methods.
From this sample, it is unclear whether FRBs with periodic components may represent a separate and rare class or whether a larger number of FRBs would show periodic behaviours if sufficiently many components were detected.

Millisecond to second periodicities may suggest that the bursts are generated by Galactic radio pulsars that have been misidentified as extragalactic.
However, FRB 20191221A has a measured DM $\sim$ 4 times larger than the maximum value expected by models of the Milky Way electron content.\cite{ne2001,ymw17}
We searched for evidence of ionized \cite{anderson2014wise,green2019revised} or star-forming\cite{avedisova2002catalog} regions in the direction of the FRB that could account for the excess DM but found none. 
We conclude that the source is likely extragalactic.

No additional bursts have been detected from FRB 20191221A up to March 10th, 2021 above an S/N of 9 at a position consistent within $\Delta\text{RA}=2.2$\,deg\,$\cos^{-1}(\text{Dec})$\, and $\Delta\text{Dec}=1$\,deg of these three sources using an algorithm built on a density-based spatial clustering of applications with noise (\texttt{DBSCAN}).\cite{d21}
The nominal DM range threshold for the clustering was set to 13~pc\,cm$^{-3}$, corresponding to the largest DM uncertainty in the real-time pipeline of CHIME/FRB.\cite{abb+18}
CHIME/FRB continues to monitor the sky location of these FRBs daily for possible additional bursts.

Multi-component bursts that are associated with repeating sources of FRBs often exhibit downward-drifting subbursts,\cite{hss+19} however, all of the components forming FRB 20191221A show a similar spectrum.
This is visible in $\sim$ 5\% of the FRB population detected by CHIME/FRB.\cite{pgk+21}
It must be noted that the spectrum of FRB 20191221A is affected by the telescope response, mainly due to the detection at a location offset from the centre of a formed beam, which produces strong bandpass effects (see Methods).\cite{abb+19a}

Our modeling of the pulse profile of FRB 20191221A, visible in Fig.~\ref{fig:waterfall} and described in detail in the Methods, shows that its single components have a relatively narrow width of 4(1)\,ms on average, even though they overlap due to an extreme scattering timescale $\tau_s=340(10)$\,ms at 600\,MHz.
Although this estimate may be affected by unresolved features in the profile mimicking an exponential decay, it is clear that the FRB emission experienced strong scattering, significantly in excess of the Galactic contribution expected given its sky position,\cite{ne2001} pointing to propagation through a turbulent plasma.
The pulse width corrected for the pulse broadening corresponds to a duty cycle of $\sim 1.8\%$, consistent with Galactic radio pulsars.\cite{mht+05}
It is worth noting that all of the emission from this FRB is consistent with single components overlapping due to the large scattering and no envelope of emission is required in our fit, whose residuals are consistent with noise, as is visible in Fig.~\ref{fig:waterfall}.

Leading theories for the origin of FRBs are related to magnetars\cite{pww+19,abb+20,brb+20}
and are divided into models where the emission is either generated in the star's magnetosphere or triggered in plasma regions by a flare of the star.
The detection of periodicity is naturally explained by the first class of magnetar models, and it has been extensively observed in Galactic neutron stars, albeit with orders of magnitude lower luminosities.\cite{mht+05,crh+06}
By contrast, the second class of models does not necessarily predict a millisecond modulation in the emitted signal.\cite{mms19}

The periodic structures in the bursts could be explained by a rotating neutron star with beamed emission similar to Galactic radio pulsars where, for an unknown reason, a train of single pulses has an abnormally high luminosity for a short period of time.
The period and jitter in the ToAs observed for FRB 20191221A are compatible with those seen in Galactic pulsars.\cite{mht+05} 
Alternatively, bright radio pulsars and magnetars sometimes show micro-structures in some of their single pulses with profiles that are similar to those of the FRBs reported on here\cite{ccd68,pmp+18} and quasi-periodic separations\cite{han71} in some cases.
However, it was shown that the ToAs of single components of FRB 20191221A do not deviate significantly from a strict periodicity.
Also, if these components are structures forming a single pulse, its width would be $\sim 4$ times larger than the widest component ever observed in a radio pulsar.\cite{mht+05}
Finally, as opposed to typical micro-structures seen in pulsars, the components do not require an envelope of emission.

Recent theories have also predicted the detection of periodicity in FRB sub-bursts if magnetar crustal oscillation frequencies can be directly related to oscillation modes on the surface of the magnetar during outburst,\cite{wc20,lkz20}.
A periodicity of 216.8\,ms corresponds to a surface oscillation frequency of 4.6\,Hz, within the range observed previously in Galactic magnetars.\cite{hdw+14}

One unlikely scenario to explain the larger radio luminosity of the FRB compared to Galactic sources is that it may represent the observation of a gravitationally micro-lensed extragalactic pulsar.
Alternatively, the single components of periodic FRBs could be generated by the magnetospheric interaction of merging neutron stars.
These possibilities are explored in the Methods, along with suggested tests of these models. 
In the meantime, CHIME/FRB is continuing to detect hundreds of FRBs, which should allow for additional periodicities to be detected in the near future.

\printbibliography[segment=\therefsegment,heading=subbibliography]

\clearpage

\clearpage
\begin{methods}
\renewcommand{\figurename}{Extended Data Figure}
\setcounter{figure}{0}
\renewcommand{\tablename}{Extended Data Table}
\setcounter{table}{0}

\section*{CHIME/FRB sensitivity beams}
The CHIME radio telescope is a transit interferometer formed by 1024 dual-polarisation antennas observing in the 400-800\,MHz range. 
The field of view of single antennas summed together incoherently is defined as the primary beam of the telescope, whose FWHM sensitivity spans $\sim 110$ degrees in the N-S direction and $1.3$ ($2.5$) degrees the E-W direction at the top (bottom) of the observing bandwidth.\cite{abb+18}
The telescope antennas can be added coherently to produce a \emph{formed} (or \emph{synthesized}) beam in one or more directions and increase the telescope sensitivity towards those directions.
Formed beams have a size between $\sim$ 0.3 and 0.7 degrees, depending on the observing frequency and zenith angle.\cite{abb+18}
Therefore, their sensitivity and bandpass vary spatially more rapidly than those of the primary beam.

In the real-time search for FRBs, 1024 beams are formed on the sky within the primary beam via an FFT (FFT beams).\cite{nvp+17}
The total intensity measured by these FFT beams as a function of time and frequency is referred to as \emph{intensity} data. 
Intensity data have a time resolution of $0.98304$\,ms and are divided into $16,384$ channels.
Additionally, for a part of the detected FRBs, $\sim 100$\,ms of \emph{baseband} complex voltages are also stored.\cite{abb+18}
The baseband data have a time resolution of $2.56$~$\mu$s, are divided into $1024$ frequency channels and contain full polarisation information.
Thanks to the phase information available in baseband data, synthesized beams can be formed to virtually any position on the sky within the telescope's field of view.\cite{mmm+21}

\begin{figure}
\begin{center}
\includegraphics[width=\textwidth]{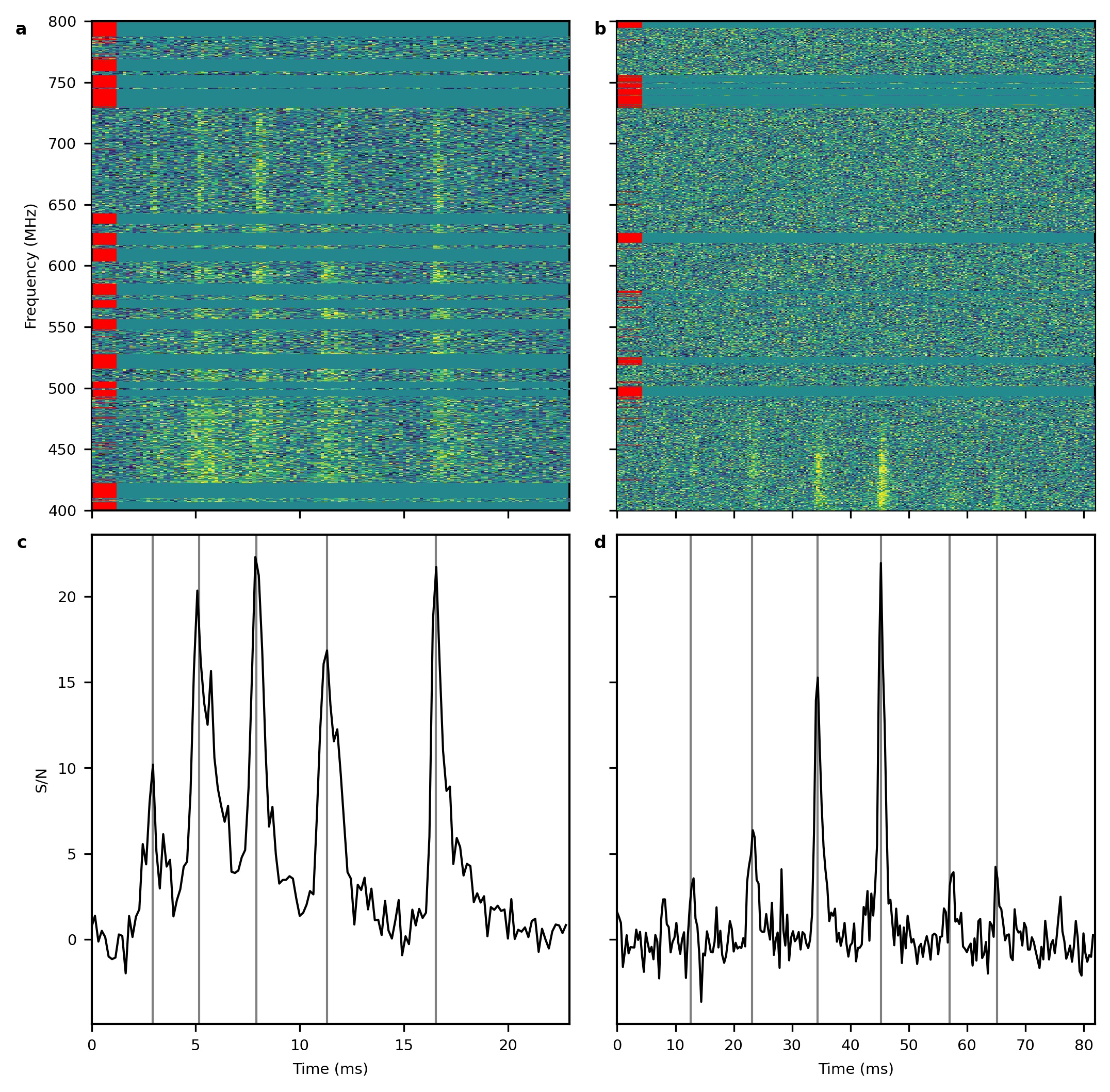}
\caption{
\textbf{Radio signal from FRBs 20210206A (a, c) and 20210213A (b, d).}
\textbf{a, b}, Waterfall plots of the signal intensity (colour-coded) as a function of time and frequency. Frequency channels missing or masked due to radio frequency interference are replaced with off-burst median values and are indicated in red. 
Effects of dispersion have been removed, and data have been plotted at the native frequency resolution of 390.625\,kHz and at a time resolution of 0.16 and 0.32\,ms, respectively.
\textbf{c, d}, in black, the pulse profiles obtained by averaging the frequency channels of the waterfall plots where signal is visible. Peak locations are highlighted by vertical lines.
}
\label{fig:waterfall_ED}
\end{center}
\end{figure}

\begin{figure}
\begin{center}
\includegraphics[width=\textwidth]{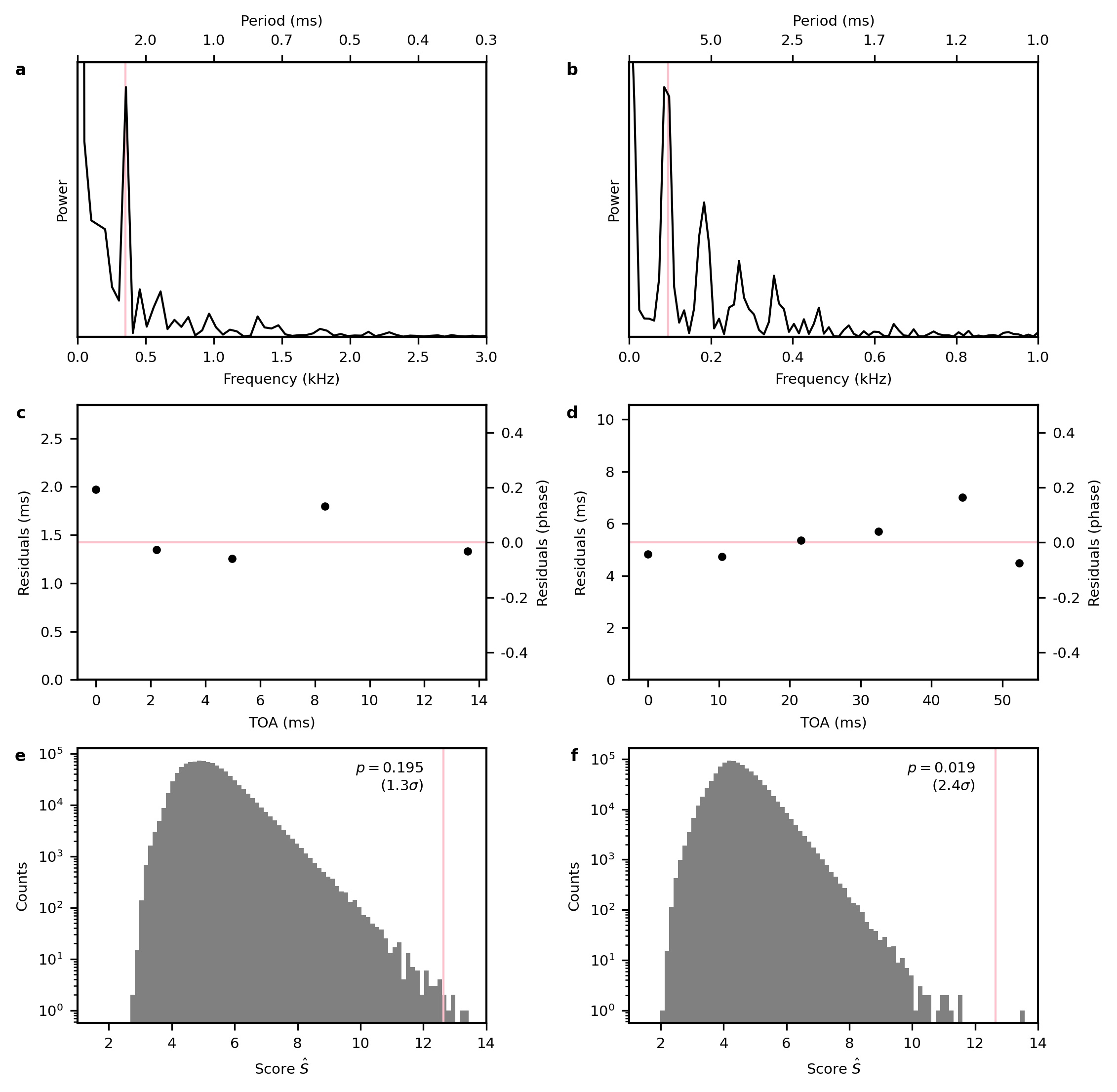}
\caption{
\textbf{Periodicity analysis of FRBs 20210206A (a, c, e) and 20210213A (b, d, f).}
\textbf{a, b}, Power spectrum obtained with a discrete Fourier Transform of the pulse profile. Vertical pink lines indicate the periods reported in Extended Data Table~\ref{tab:bursts_ED}.
\textbf{c, d}, Residuals of a timing analysis assuming that the peaks forming the FRB profile are separated by integer numbers times these periods, respectively. 1-$\sigma$ error bars are often smaller than the symbol sizes. Horizontal pink lines indicate a phase of zero around which residuals have been rotated.
\textbf{e, f}, Study of the statistical significance of the measured periodicity by using the periodicity-sensitive score $\hat S$. The grey histograms have been obtained with an ensemble of simulations, whereas the value measured for each FRB is represented with a vertical pink line. The corresponding probability of obtaining such a periodicity by chance is indicated on the plots.
}
\label{fig:periodicity_ED}
\end{center}
\end{figure}

\section*{Sources of interest}
We initially identified FRB 20191221A as an interesting source due to the number of peaks in its pulse profile. Only intensity data was stored for this source. 
We then visually inspected the sample of events detected by CHIME/FRB for other similar sources. 
We found none with comparable characteristics and the closest for number of components were found to be FRBs 20210206A and 20210213A (Extended Data Fig.~\ref{fig:waterfall_ED}).
We repeated the timing analysis previously performed for FRB 20191221A.
The results are summarised in Extended Data Fig.~\ref{fig:periodicity_ED}.
FRBs 20210206A is formed by 5 separate components whose inferred period of $2.8(1)$ ms has a spurious detection probability as high as $0.2$, making it uncertain at best.
FRB 20210206A is formed by 6 separate components whose inferred period of $10.7(1)$ ms has a spurious detection probability of $0.02$, higher than the previous source but still not conclusive.
For this reason, and given the large jitter of the timing residuals visible in Extended Data Fig.~\ref{fig:periodicity_ED}, we consider the periodicities detected in these two bursts to be only suggestive and we did not use them in our analysis.
However, these FRBs demonstrate the existence of a broad distribution in the number of components in burst profiles, with some sources showing possible periodic separations.
Hints of shorter periodic separations of $\sim 0.3$, $\sim 2.3$, and $\sim 415$\,$\mu$s in the components of other FRBs have also been recently presented.\cite{nhk+21b,mpp+21,pvb+22}

Periodicities of 16.3 and 157 days in the activity levels of two FRBs have been previously reported.\cite{aab+20,rms+20} 
However, these periodicities do not represent an actual time delay between subsequent bursts.
This, and the very different timescales, argue that the two phenomena are unrelated.

\begin{table}
\begin{center}
\begin{tabular}{lcc}
\hline
Parameter & 20210206A & 20210213A \\
\hline
MJD & 59251.11021004 & 59258.46924637 \\
RA J2000 (deg) & 53.86(2) & 192.6(1) \\
Dec J2000 (deg) & 52.743(7) & 83.28(2) \\
$l$ (deg) & 146.359(9) & 122.97(2) \\
$b$ (deg) & $-$2.503(8) & 33.85(2) \\
DM (pc\,cm$^{-3}$) & 361.35(7) & 482.5(2) \\
Period (ms) & 2.8(1) & 10.7(1) \\
Period significance ($\sigma$) & 1.3 & 2.4 \\
Average width (ms) & 0.068(5) & 0.42(4) \\
Scattering (ms) & 1.25(2) & 0.78(5) \\
Fluence (Jy ms) & 47(14) & 8.4(2.9) \\
Peak Flux (Jy) & 5.7(1.8) & 1.2(5) \\
Exposure$^*$ (hours) & 99.2(1) & 196(4), 496.7(2)  \\
\hline
\end{tabular}
\end{center}
$^*$ For circumpolar sources ($\delta > +70^\circ$), the two entries correspond to exposure in the upper and lower transit, respectively.
\caption{
{\bf Properties of FRBs 20210206A and 20210213A.} 
  All quantities have been calculated with baseband data except for fluences and fluxes.
  Uncertainties are reported at 1-$\sigma$ confidence level.  The arrival time is that of the brightest sub-burst at the Solar System's barycentre and infinite frequency. The DM is calculated to maximize the peak S/N in the timeseries. The scattering timescale is referenced to the centre of the band, i.e. $\sim 600$\,MHz.
  Fluence is for the full band-averaged profile, and peak flux is the maximum in the profile (with 1-ms time resolution).
\label{tab:bursts_ED}
}
\end{table}

Both intensity and baseband data are available for FRBs 20210206A and 20210213A. 
This allowed us to localize the sources with sufficient precision to form beams in these directions and, therefore, limit the effect of their bandpass.\cite{mmm+21}
Therefore, the lack of emission above $\sim 500$\,MHz seen for FRB 20210213A is astrophysical.
The DM of FRB 20210213A is $\sim$ 10 times the expected Galactic contribution corresponding to an estimated redshift of $z\sim0.4$.\cite{bkm21}
FRB 20210206A, located at a Galactic latitude $b=-2.5$\,deg, has a DM $\sim$ 1.5 times larger than the Galactic contribution, placing it at a supposed redshift $z\sim0.1$ but with the caveat that its extragalactic nature is less certain given the larger uncertainty of models at low latitudes.\cite{hww+21}
As opposed to FRB 20191221A, FRBs 20210206A and 20210213A show narrower widths and shorter scattering timescales (see Extended Data Table~\ref{tab:bursts_ED}) typical of the FRB population.\cite{pgk+21}
The baseband data also allowed us to study the polarisation properties of FRBs 20210206A and 20210213A.
A rotation measure $\mathrm{RM} = +193.6(1)$\,rad\,m$^{-2}$ has been measured for FRB 20210206A, suggesting a significant extragalactic contribution. 
On the other hand, FRB 20210213A appears to be unpolarised. Possible reasons for this are discussed in the following, together with a detailed description of the polarisation analysis.

\section*{Properties of the bursts}
To localize an FRB with CHIME/FRB intensity data, we fit the spectra of the burst detected in different FFT beams with a model of the CHIME/FRB beams and an underlying burst spectrum using a Markov chain Monte Carlo (MCMC) method.\cite{fhlg13}
The model of the CHIME/FRB beams contains a description of both the synthesized\cite{nvp+17,msn+19} and primary\cite{1stcatalog} beams and the underlying spectrum is modeled as a Gaussian. 
The free parameters are therefore width, mean, and amplitude of the underlying Gaussian model spectrum, along with the sky position. 
We use a flat prior on the position of the event that spans $5^{\circ}$ to either side of meridian in E-W, while in N-S the prior spans the extent of the beams that detected the event. 
The position and uncertainties are derived from the 2D posterior distribution (in `x', the E-W coordinate, and `y', the N-S coordinate) marginalized over the parameters of the Gaussian spectral model. Since FRB 20191221A did not have baseband data available, the position reported in Table~\ref{tab:bursts} is derived from the intensity localization described in the present section. The posterior probability distribution is double-peaked in the E-W direction and so two positions are reported for the event.

A detailed description of the algorithm used to obtain the sky position of FRBs by using baseband data has been presented elsewhere.\cite{mmm+21}
In summary, a grid of partially overlapping beams is produced around the intensity localization and a total S/N value is calculated in each beam.
The resulting intensity map of the signal is fitted with a mathematical model describing the telescope response to determine the source position. 
The localization and its uncertainties have been calibrated with a sample of sources with a known position to account for any unmodelled systematics.\cite{mmm+21}

Flux and fluence calculations are determined using the intensity data for each burst with the same method presented in previous CHIME/FRB papers. \cite{abb+19a, abb+19b, jcf+19, abb+19c, fab+20, aab+20} In summary, meridian transits of steady sources with known spectra are used to sample the conversion between beamformer units and Janskys as a function of frequency across the N-S extent of the primary beam. For each burst, the beamformer to Jansky conversion closest in zenith angle (assuming N-S symmetry) is applied to the intensity data to obtain a dynamic spectrum in physical units roughly corrected for N-S primary beam variations. Fluence values are obtained from integrating the burst extent in the band-averaged time series, and peak flux values are taken to be the maximum value within the burst extent (at 0.98304 ms resolution). Uncertainties are estimated using steady source observations.
The calibration procedure described above does not correct for burst attenuation due to the synthesized beam pattern and E-W primary beam profile. Fluences and fluxes derived from this method are best interpreted as lower limits, with an uncertainty on the limiting value. This is what we report in Table~\ref{tab:bursts} for FRB 20191221A. However, for bursts that have a baseband localization, we can achieve more realistic fluence results by using the beam model to scale between the location of the calibrator at the time of transit and the location of each FRB. This is what we report in Extended Data Table~\ref{tab:bursts_ED} for FRBs 20210206A and 20210213A.

The exposure of the CHIME/FRB system to the sources reported in this work was determined for the interval from August 28, 2018 to March 1, 2021. For each source, the exposure is calculated by summing the duration of daily transits across the FWHM region of the synthesized beams at 600 MHz. Two of the three sources, having declinations $> +70^\circ$, transit through the primary beam twice per day. These sources have their upper and lower transit exposures calculated separately as the beam response for the two transits is different. 
While calculating the total exposure, we include daily transits for which the CHIME/FRB detection pipeline was fully operational, which is determined using recorded system metrics. Transits occurring on days when the detection pipeline was being tested or upgraded are not included. Additionally, system sensitivity varies on a day-to-day basis due to daily gain calibration as well as changes in the detection pipeline and the RFI environment. We evaluate the variation in sensitivity for each sidereal day using observations of 120 Galactic pulsars with the CHIME/FRB system.\cite{jcf+19, fab+20} For each source, transits for which the sensitivity varied by more than 10\% from the median in the aforementioned observing interval are excised from the total exposure. On average, the observing time corresponding to the excised transits amounted to 4\% of the exposure for each source.
The uncertainty in the source declination is a source of error in the measurement of the exposure. The source declination dictates where the transit path cuts across a synthesized beam with the transit duration being maximum if the path crosses the beam centre and zero if the path lies between two beams.\cite{abb+19c} We estimate the resulting uncertainty in the exposure by generating a uniform grid of positions within the 68\% confidence localization region for each source. The reported exposure in Table~\ref{tab:bursts} and Extended Data Table~\ref{tab:bursts_ED} is the average for these sky positions with the error corresponding to the standard deviation.

\section*{Modeling of pulse profiles}
FRB 20191221A is composed of multiple peaks overlapping due to a large broadening caused by scattering. To properly calculate the periodic separation among its components, it is important to avoid human bias in selecting significant peaks in the pulse profile.
We reduced human bias in the following way.
First, we smoothed the pulse profile using the Savitzky-Golay filter as implemented in \texttt{SciPy}.\footnote{\url{https://scipy.org/}}
The filter requires two input parameters, the window length, and the polynomial order. We explored the space of the two parameters up to a window of 600\,ms and a polynomial order of 12.
For each combination, separate peaks in the smoothed pulse profiles were identified from local maxima if the peaks were wider than 3 bins to avoid noise spikes.

\begin{figure}
\begin{center}
\includegraphics[width=.5\textwidth]{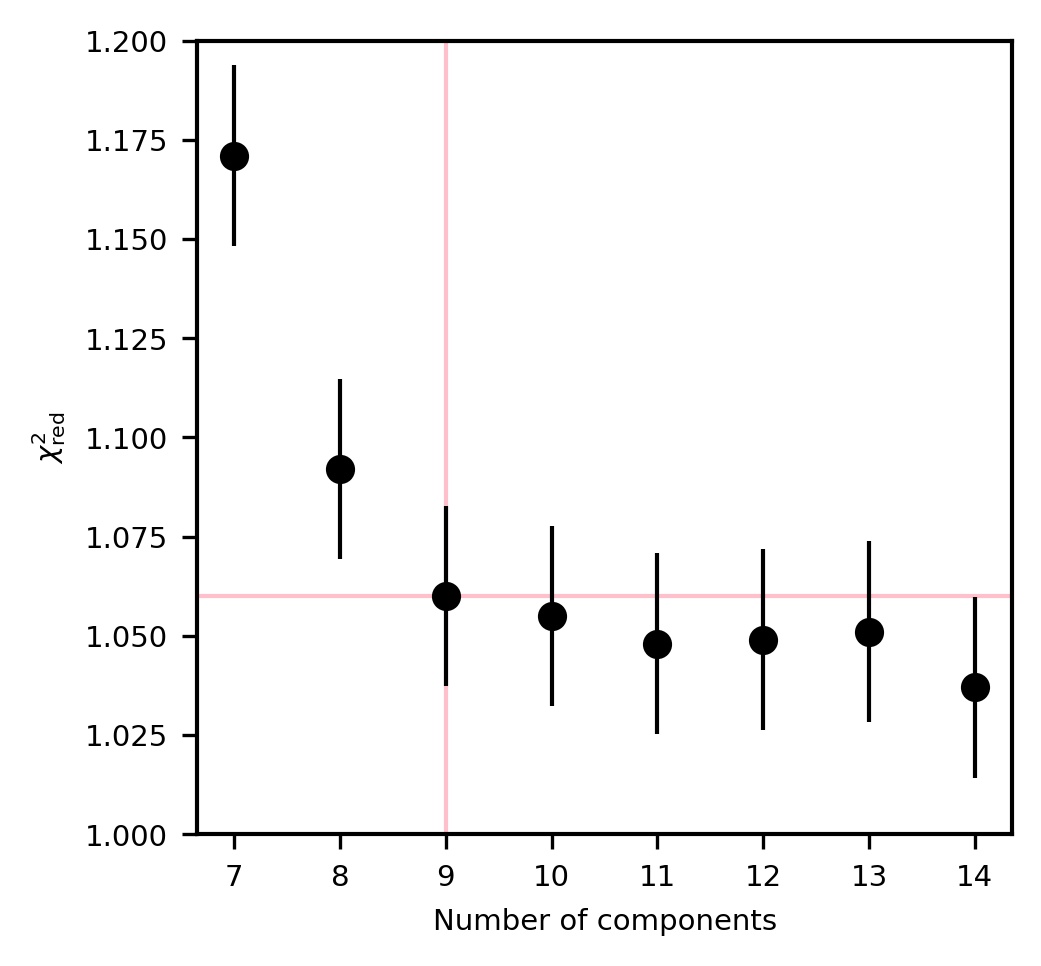}
\caption{
\textbf{Reduced chi-square test as a function of the number of components used to model the profile of FRB 20191221A.}
The vertical line highlights the chosen number of components, while the horizontal line is placed at the $\chi^2_\textrm{red}$ value for 9 components.
The minimum $\chi^2_\textrm{red}$ variation that can be measured confidently with our data is estimated with Eq.~\ref{eq:chi2_err} and plotted as en error bar for each number of peaks.
}
\label{fig:657chivspeaks}
\end{center}
\end{figure}

We grouped together different combinations of parameters yielding the same number of peaks, noting their location in the profile and excluding values lower than 7 and higher than 14 peaks as visually implausible.
For each value of the peak number between 7 and 14, the different peak locations resulting from each of the initial parameter combinations were grouped with a kernel density estimator, varying its window until we obtain the expected number of peaks. 
We apply this method to obtain seven combinations of initial peak positions composed of 7 to 14 peaks.
We model each peak using the exponentially modified Gaussian (EMG) described by\cite{mck14,1stcatalog}
\begin{equation}
\label{eq:EMG}
    f(t; \mu, w, \tau) = 
    \frac{A}{2\tau} 
    \exp\left(\frac{\mu - t}{\tau}  + \frac{2\sigma^2}{\tau^2}\right)
    \text{erfc} \left( \frac{(\mu - t)\tau + w^2}{w\tau\sqrt{2}}\right),
\end{equation}
where erfc is the complementary error function, $A$ is the signal amplitude, $\mu$ and $w$ are the Gaussian mean and width, respectively, and $\tau$ is the scattering timescale. 
For each number of peaks, the pulse profile is modelled with a sum of one EMG function per peak through an MCMC sampling using \texttt{emcee}\cite{fhlg13} with wide uniform priors for all parameters.
The scattering timescale is set to be the same for all peaks, while the other parameters are allowed to vary.
The resulting $\chi^2_\text{red}$ values, presented in Extended Data Figure~\ref{fig:657chivspeaks}, are used to select the number of peaks yielding the best model.
We estimate the minimum variation in the $\chi^2_\text{red}$ value that we can measure confidently given the number of free parameters $N$ as
\begin{equation}\label{eq:chi2_err}
    \sigma = \sqrt{\frac{2}{N}}.
\end{equation}
As visible in Extended Data Figure~\ref{fig:657chivspeaks}, the model with 9 components has the smallest $\chi^2_\text{red}$ that deviates significantly from the previous values.
Therefore, we choose this as the best model to reproduce the data and use its parameters in our analysis.
Choosing a different number of peaks in the profile leads to values of the significance of the periodicity that are still very high, especially for a number of peaks between 9 and 12.

\begin{table}
\begin{center}
\begin{tabular}{cccc}  
\hline
Component & 20191221A & 20210206A & 20210213A \\ 
\hline
1 & 0(6)         & 0.00(1)   & 0.00(5)  \\
2 & 430(9)       & 2.221(5)  & 10.53(8) \\ 
3 & 652(3)       & 4.974(6)  & 21.70(3) \\ 
4 & 1086.2(8)    & 8.358(8)  & 32.60(2) \\ 
5 & 1520(2)      & 13.580(9) & 44.4(1)  \\
6 & 1736(1)      & $\cdots$  & 52.48(5) \\ 
7 & 1952(1)      & $\cdots$  & $\cdots$ \\
8 & 2171(2)      & $\cdots$  & $\cdots$ \\ 
9 & 2604(2)      & $\cdots$  & $\cdots$ \\
\hline
\end{tabular}
\end{center}
\caption{
\textbf{List of times of arrival (ToAs) for the peaks forming each event.}
The ToA of each component is reported in milliseconds relative to the first peak.
1-$\sigma$ uncertainties on the last digit are indicated in parenthesis.
}
\label{tab:toa_list}
\end{table}

For FRBs 20210206A and 20210213A, the components forming these two events do not overlap. Therefore, it was not necessary to perform the method described above. 
Instead, we directly used the locations of the peaks in the smoothed profiles as initial conditions for the MCMC sampling using the same model described above.
The scattering timescale and average width of the resulting models are presented in Table~\ref{tab:bursts} and Extended Data Table~\ref{tab:bursts_ED}, while the peak positions (or ToAs) relative to the first one in each profile are reported in Extended Data Table~\ref{tab:toa_list}.

\section*{Timing analysis}
The periodicities of the three FRBs were initially investigated through a power spectrum of the pulse profile.
This has been computed as the absolute square of the Fast Fourier Transform (FFT) of the pulse profiles.
The FFT has been calculated with the implementation offered by the \texttt{SciPy} module. 
The resulting power spectra presented in Fig.~\ref{fig:periodicity} and Extended Data Fig.~\ref{fig:periodicity_ED} show clear peaks for the three sources.
The periods corresponding to the most prominent peaks in the power spectra have been refined through an additional timing analysis.
We ran a least-squares fit as implemented in \texttt{SciPy} using a simple model with the period and an arbitrary overall phase as the only free parameters, calculating residuals as the modulo of ToAs (reported in Extended Data Table~\ref{tab:toa_list}) and the period.
The results of these fits are the values reported in Table~\ref{tab:bursts} and Extended Data Table~\ref{tab:bursts_ED}.

\begin{table}
\begin{center}
\small
\begin{tabular}{lccccccccc}
\hline
FRB & ToAs & Gaps & Trials & $p$-value (${\hat S})$ & $\sigma$ (${\hat S}$) & $P$ (ms) & $Z_{1}^{2}$ & $p$-value (${Z_{1}^{2}}$) & $\sigma$ (${Z_{1}^{2}}$) \\ 
\hline
20191221A & 9 & 3 & 1365 & 6.7\,$\times$\,10$^{-11}$ & 6.5 & 217.3 & 18.0 & 5.0\,$\times$\,10$^{-11}$ & 6.22 \\ 
20210206A & 5 & 0 & 6 & 0.195 & 1.3 & 2.8 & 6.9 & 0.9998 & 0.0002 \\ 
20210213A & 6 & 1 & 15 & 0.019 & 2.4 & 10.8 & 9.7 & 0.57 & 0.58 \\ 
\hline
\end{tabular}
\end{center}
\caption{
\textbf{Statistical significance of the FRB periodicities.}
For each FRB, we report the number of ToAs measured from the profile, the number of gaps and trials considered in the $\hat{S}$-periodicity analysis, and the resulting probability and significance.
The values in the last four columns are derived using the Rayleigh ($Z_{1}^{2}$) test and show the period obtained in the analysis, the resulting value of the test, and the false alarm probability and significance of the periodicities.
}
\label{tab:p_values}
\end{table}

\section*{Significance of the periodicity: $\hat S$ score}
The following steps were used to estimate the significance of the periodicity calculated for each of the three sources and reported in Table~\ref{tab:bursts} and Extended Data Table~\ref{tab:bursts_ED}.
\begin{enumerate}
\item For each event, we use the ToAs reported in Extended Data Table~\ref{tab:toa_list} to compute a statistic $\hat S$ which is
sensitive to periodicity. The construction of $\hat S$ is described below. 
\item To assign a statistical significance, we use a frequentist approach.
We evaluate the statistic $\hat S$ on simulated events and rank the
`data' value $\hat S_{\rm data}$ within the ensemble of simulated
values $\hat S_{\rm sim}$, obtaining a $p$-value.
The results are summarized in Extended Data Table~\ref{tab:p_values}.
\item For FRB 20191221A, there is an extra step. With $10^8$ simulations,
we find that none of the $\hat S_{\rm sim}$ values exceed $\hat S_{\rm data}$.
This shows directly that the level of periodicity in this 9-component event is extraordinarily unlikely to occur by random chance.
To assign a $p$-value, we fit an analytic model PDF to the tail of the $\hat S_{\rm sim}$ distribution and integrate the model PDF.
\end{enumerate}
In the rest of this section, we describe the details in the
above steps.

\begin{figure}
\begin{center}
\includegraphics[width=.5\textwidth]{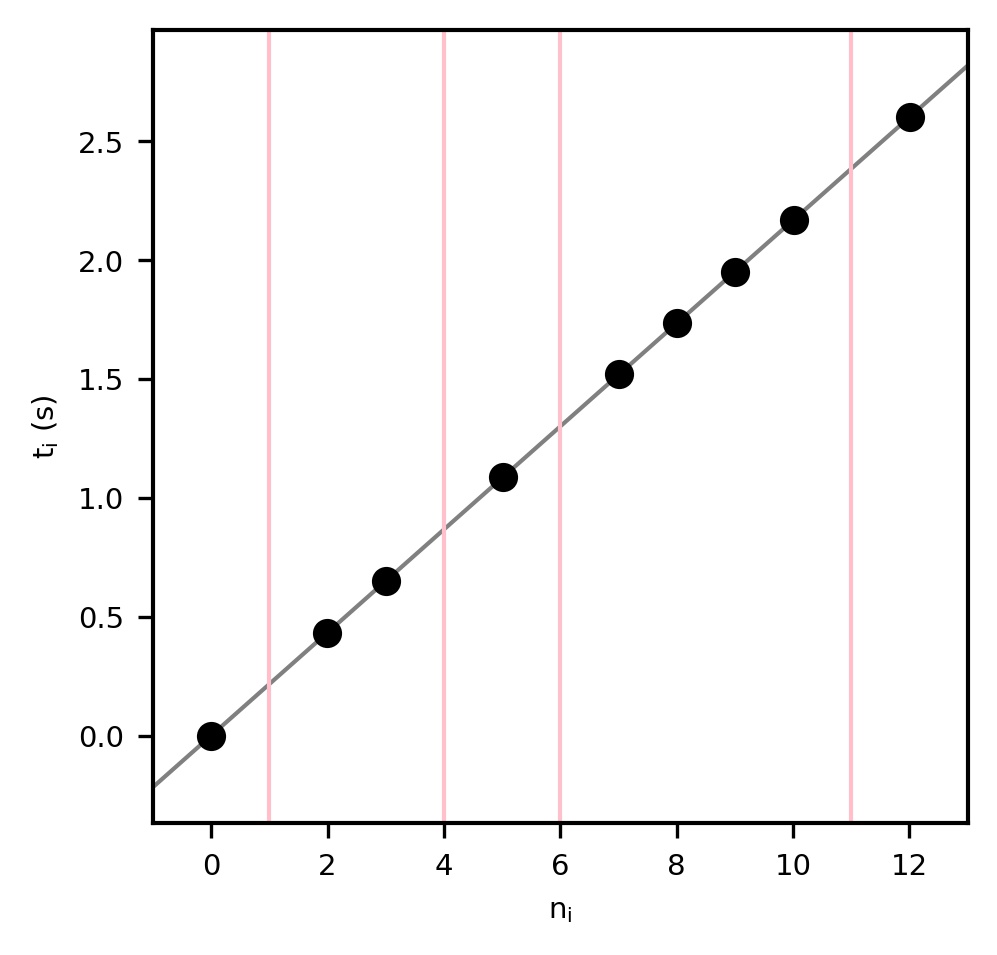}
\caption{
\textbf{Times of arrival (ToAs) of the components of FRB 20191221A as a function of their measured cycle.}
The cycle is defined in Eq.~\ref{eq:nj12}. 
The periodicity appears clearly as the points fall nearly on the straight gray line, which highlights the trend expected for a period of $216.8$\,ms.
Vertical lines mark gaps where no pulse is observed within one period.
}
\label{fig:nt_regression}
\end{center}
\end{figure}

To motivate the definitions which follow, consider Extended Data Figure~\ref{fig:nt_regression}. 
Each point is one pulse in the 9-component event, with the ToA $t_i$ on the $y$-axis. The values on the $x$-axis are the 9-element integer-valued vector 
\begin{equation}
n_i = (0, 2, 3, 5, 7, 8, 9, 10, 12)  \label{eq:nj12}
\end{equation}
Periodicity appears in Extended Data Figure~\ref{fig:nt_regression} as the points falling nearly on a straight line. 
Note that there are four `gaps' in this example, i.e.~pulse periods where no pulse is observed (either because it is physically absent or buried in the noise).
Each gap is represented by consecutive entries in $n_i$ which differ by 2 (rather than 1).
Based on this picture, we construct statistics as follows.
For a fixed choice of gap vector $n_i$, we fit the points $(n_i,t_i)$ in Extended Data Figure~\ref{fig:nt_regression} to a straight line 
\begin{equation}
t_i = f^{-1} n_i + T_0 + r_i  \label{eq:nt_regression}
\end{equation}
where this equation defines the residuals $r_i$, and the parameters $(f,T_0)$ have been chosen to minimize $\sum_i r_i^2$.
Note that this fitting procedure weights all ToAs equally, and does not use statistical errors on ToAs. We define a statistic
\begin{equation}
\hat L[n] = \frac{1}{2} \log\left( \sum_i \frac{ (t_i-\bar t_i)^2}{r_i^2} \right)
\end{equation}
By construction, $\hat L[n]$ measures the extent to which the points fall on a straight line for a fixed choice of gap vector $n_i$.
We define the periodicity-sensitive statistic $\hat S$ by trying all possible values of $n_i$
\begin{equation}
\hat S = \max_n\left( \hat L[n] \right)
\end{equation}
The maximum in the equation is taken over all trial gap vectors $n_i$ with $\le G$ gaps, where $G$ is an input parameter to the pipeline.
More formally, we take the maximum over integer-valued vectors $n_i$ such that $n_i=0$, $n_{i-1} < n_i$, and $n_p < n_1 + p + G$, where $p=\mbox{len}(n)$ is the number of peaks.
The number of such vectors is
\begin{equation}
N_{\rm trials} = \frac{(G+p-1)!}{G! \, (p-1)!}  \label{eq:ntrials}
\end{equation}
By default, we choose $G = N_{\rm gaps}+1$, where $N_{\rm gaps}$ is the number of gaps that are empirically seen in each event. This is a conservative choice, which deliberately ensures that $N_{\rm trials}$ is a few times larger than the value obtained with $G=N_{\rm gaps}$ (see Extended Data Table~\ref{tab:p_values}).

Our procedure for simulating ToAs has two parameters: a mean spacing $\bar d$, and a dimensionless `exclusion' parameter $0 \le \chi \le 1$ defined by
\begin{equation}
\chi = \cfrac{\mbox{Minimum allowed spacing between pulses}}{\mbox{Mean spacing $\bar d$ between pulses}}.
\label{eq:chi_def}
\end{equation}
We simulate a sequence $t_1 < \cdots < t_p$ of ToAs by independently randomly generating arrival time differences $d_i = (t_i - t_{i-1})$ from the uniform probability distribution $p(d)$ defined by
\begin{equation}
p(d) = \left\{ \begin{array}{cc}
 1/(2\bar d-2\chi\bar d) & \mbox{if } \chi \bar d \le d \le (2-\chi) \bar d \\
  0 & \mbox{otherwise} \end{array}.
\right.
\label{eq:toa_diff_dist}
\end{equation}
We also tried an exponential distribution, but find that it gives lower $p$-values (higher significance). To be conservative, we use the uniform distribution throughout.
When we assign a statistical significance by ranking $\hat S_{\rm data}$ within a histogram of $\hat S_{\rm sim}$ values, we find that the value of $\bar d$ does not affect the statistical significance, while the statistical significance decreases as $\chi$ is increased in
the simulations. 
However, the statistical significance is not strongly dependent on $\chi$ for a reasonable range of the parameter. 
Therefore, we choose $\chi=0.2$ to represent our analysis.

For FRB 20191221A, there is an extra step. With $10^8$ simulations, we find that none of the $\hat S_{\rm sim}$ values exceed $\hat S_{\rm data}$.
Therefore, we fit the $10^5$ largest values $S_i$ of the $\hat S_{\rm sim}$-distribution to an analytic PDF of the form
\begin{equation}
p(S) \propto e^{-aS^q}
\end{equation}
by maximizing the likelihood $\prod_i P(S_i|a,q)$.
The analytic PDF $p(S)$ is an excellent visual fit to the tail of the $\hat S_{\rm sim}$ distribution.
More quantitatively, a KS test shows no statistical difference between $p(S)$ and the simulations ($p$-value 0.58).
To assign a bottom-line $p$-value, we integrate the analytic PDF from $S=\hat S_{\rm data}$ to $S=\infty$.
This gives a $p$-value of $6.7 \times 10^{-11}$, corresponding to Gaussian significance 6.5$\sigma$.

The preceding derivation of the statistic $\hat S$ was heuristic, based on intuition from Extended Data Figure~\ref{fig:nt_regression}.
However, $\hat S$ can also be interpreted as a likelihood ratio statistic.
This provides a systematic derivation, and also shows that $\hat S$ is near-optimal.
Let $H_0$ be the null hypothesis that the ToAs $t_i$ are Gaussian distributed with mean $T$ and variance $\sigma^2$. 
In this model, the conditional likelihood of obtaining ToAs $t_i$ given model parameters $(T,\sigma)$ is
\begin{equation}
P(t_i|H_0,T,\sigma) = \prod_i \frac{1}{\sqrt{2\pi\sigma^2}}
  \exp\left(-\frac{(t_i-T)^2}{2\sigma^2} \right)
\end{equation}
Let $H_1$ be the alternate hypothesis that the $t_i$ are given by the linear regression in Eq.~(\ref{eq:nt_regression}), where the residuals $r_i$ are Gaussian with variance $\sigma^2$.
The conditional likelihood of obtaining $t_i$ given model parameters $(n_i,f,T,\sigma)$ is
\begin{equation}
P(t_i|H_1,n_i,f,T,\sigma) = \prod_i \frac{1}{\sqrt{2\pi\sigma^2}}
  \exp\left(-\frac{(t_i-T-f^{-1}n_i)^2}{2\sigma^2} \right)
\end{equation}
Then a short calculation shows that the $\hat S$ statistic is the log-likelihood ratio of the two models, after maximizing all model parameters
\begin{equation}
\hat S = \left( \max_{n,f,t,\sigma} \log P(t_i|H_1,n_i,f,T,\sigma) \right)
 - \left( \max_{n,f,t,\sigma} \log P(t_i|H_0,T,\sigma) \right)
\end{equation}

As a further test of our pipeline, we applied the $\hat S$ statistic to CHIME/Pulsar\cite{abb+21} observations of the bright pulsar PSR B1919+21. If we select three pulses with one gap, periodicity is detected at the $\sim 3\sigma$ level. If we select four pulses or more with either one or two gaps, periodicity is detected at $>4\sigma$.

\section*{Rayleigh ($Z_1^{2}$) periodicity analysis.} 
We have shown in the previous section that the $\hat S$ statistic is nearly optimal for determining the statistical significance of the periodicity observed in each FRB. However, we also calculated the significance through a different statistic that is commonly used in studies of periodicities of high-energy pulsars, the Rayleigh ($Z_1^{2}$) statistic.\cite{bbb+1983, deJager1994}
In general, the $Z_{n}^2$ test statistic is defined as
\begin{equation}
Z_{n}^{2}=\frac{2}{N}\sum_{k=1}^{n}\left[\left(\sum_{j=1}^{N}\cos k\phi_{j}\right)^{2}+\left(\sum_{j=1}^{N}\sin k\phi_{j}\right)^{2}\right],
\label{eq:Zn2}
\end{equation}
where $N$ is the number of peaks, $n$ is the number of harmonics, and $\phi_{j}$ is the phase of each ToA, $t_{j}$. The phase of each ToA is determined using $\phi_{j}$\,$=$\,$\nu t_{j}$, where $\nu$ is the modulation frequency. In the following analysis, we use $n$\,$=$\,1 harmonics in Eq.~(\ref{eq:Zn2}), which corresponds to the Rayleigh test, and the ToAs listed in Extended Data Table~\ref{tab:toa_list}. All of the ToAs were weighted equally in our analysis.

We performed a blind search for periodicity using the $Z_{1}^2$ test statistic defined in Eq.~(\ref{eq:Zn2}) and the ToAs listed in Extended Data Table~\ref{tab:toa_list}. The number of frequency trials used to search for periodicity was determined by the time resolution ($\Delta t$) and the duration ($T$) of the data containing each of the bursts. The time resolution and duration of the data used to perform the Rayleigh test were $\Delta t$\,$=$\,7.86432\,ms and $T$\,$=$\,4.215\,s for FRB 20191221A, $\Delta t$\,$=$\,81.92\,$\mu$s and $T$\,$=$\,19.6608\,ms for FRB 20210206A, and $\Delta t$\,$=$\,327.68\,$\mu$s and $T$\,$=$\,81.92\,ms for FRB 20210213A, respectively.
We searched for evidence of periodicity by calculating the $Z_{1}^2$ test statistics at a range of trial frequencies $\nu$\,$\in$\,$[\Delta \nu, \nu_{\text{nyq}}]$, where $\Delta \nu$\,$=$\,$1/T$ is the nominal frequency resolution of each data set, $\nu_{\text{nyq}}$\,$=$\,$\nu_{\text{samp}}$/2\,$=$\,1/(2$\Delta t$) is the Nyquist frequency, and $\nu_{\text{samp}}$ is the sampling frequency of the data. In addition, we oversample the frequency grid by a factor of $\mathcal{O}$\,$=$\,5. The results of this calculation are presented in Extended Data Table~\ref{tab:p_values}.

The values of $Z_1^{2}$ for the three events were converted to significance values by randomly generating arrival time differences using Monte Carlo simulations. Due to the high significance of the periodicity in FRB 20191221A, we used $\mathcal{N}_{\text{sim}}$\,$=$\,10$^{\text{10}}$ Monte Carlo simulations in our analysis of this event. For the other two events (FRBs 20210206A and 20210213A), $\mathcal{N}_{\text{sim}}$\,$=$\,10$^{\text{9}}$ Monte Carlo simulations were used to determine the significances. Using the measured ToAs, we construct random realizations of arrival time differences by drawing from a uniform probability distribution defined by Eq.~(\ref{eq:toa_diff_dist}). For each event, we perform separate sets of simulations for exclusion parameters 0\,$\leq$\,$\chi$\,$\leq$\,0.5, defined according to Eq.~(\ref{eq:chi_def}). The value of $\chi$ determines the time separation between the simulated ToAs. In the limit $\chi$\,$\rightarrow$\,1, the simulated ToAs become perfectly periodic, so we restrict $\chi$\,$\leq$\,0.5. These values of $\chi$ are used to impose a minimum time separation between ToAs in the simulations. The statistical significance of the periodicities is not strongly affected by the choice of these values of $\chi$, so we select $\chi$\,$=$\,0.2 in this analysis. We compare the distribution of maximum $Z_{1}^{2}$ test statistics obtained from each set of simulations, for a given value of $\chi$, to the $Z_{1}^{2}$ value obtained using the measured ToAs. The false alarm probability~(FAP) and equivalent Gaussian significance are calculated using
\begin{equation}
    P_{\text{FAP}} = 1 - \text{CDF}(Z_{1}^{2}) = 1-\left[1-P\left(\left.Z_{1}^{2}>Z_{1,\nu_{0}}^{2}\right|\nu=\nu_{0}\right)\right],
    \label{eq:mc_fap}
\end{equation}
where $\nu_{0}$ is the putative periodicity determined from the ToAs measured from each event.

In this analysis, the tail-fitting procedure described above is not used. Instead, the FAPs are calculated based on the number of Monte Carlo simulations that have a maximum $Z_{1}^{2}$ test statistic which exceeds the $Z_{1}^{2}$ value obtained using the measured ToAs. 
We find that the periodicity observed from FRB~20191221A has a significance of 6.2$\sigma$ using this method. The equivalent significance of the periodicities observed from FRBs 20210206A and 20210213A are both $<$\,1$\sigma$ using the Rayleigh ($Z_{1}^{2}$) statistic (see Extended Data Table~\ref{tab:p_values}).

\section*{Polarisation analysis}\label{sec:polarization}
Full polarisation information is stored for FRBs 20210206A and 20210213A.
The polarisation analysis follows a similar procedure to that previously applied to other CHIME-detected FRBs.\cite{fab+20,bgk+21} In particular, an initial RM estimate is made by applying RM-synthesis\cite{b66,bb05} to the Stokes $Q$ and $U$ data of each burst. The Stokes spectrum is extracted by integrating the polarised signal over the burst duration, where time and frequency limits have been manually adjusted to optimize the RM detection. 

\begin{figure}
\begin{center}
\includegraphics[width=.7\textwidth]{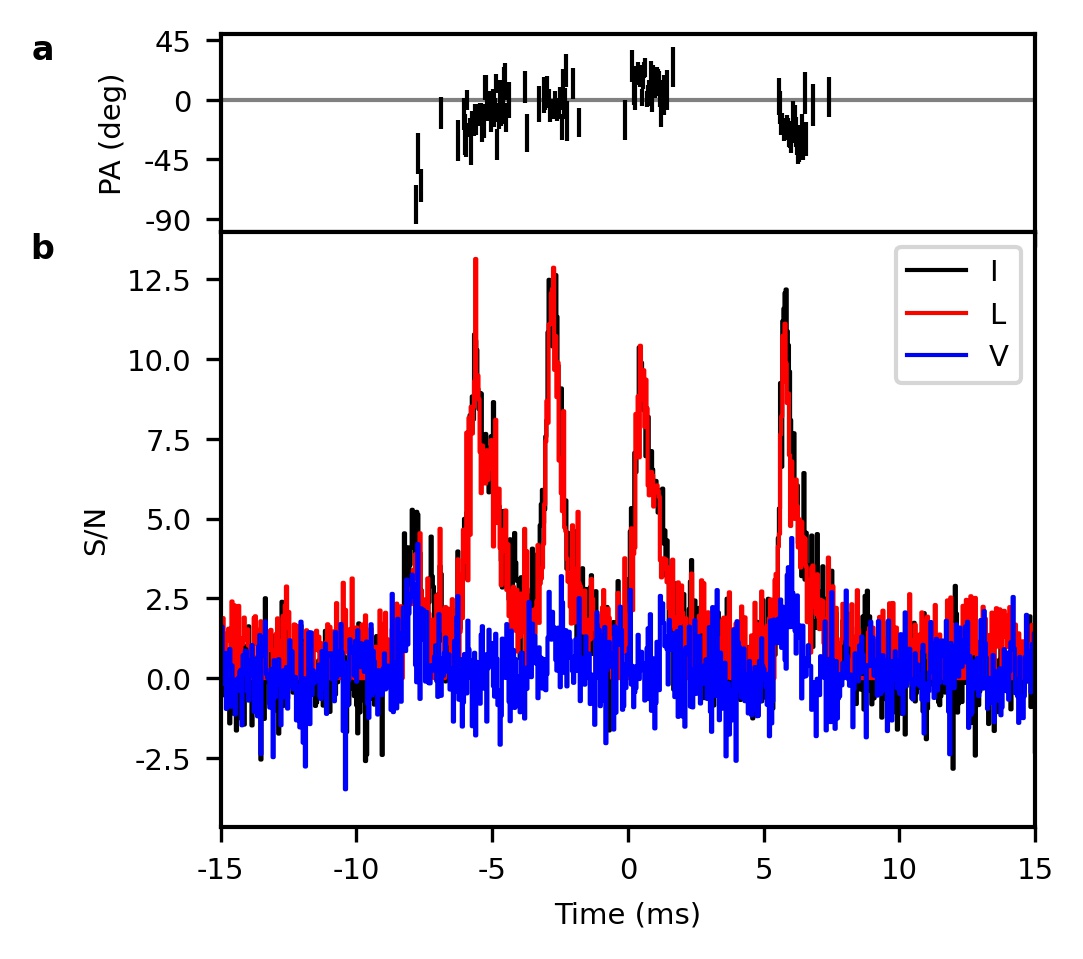}
\caption{
\textbf{Polarisation profiles of FRB 20210206A.}
\textbf{a}, The polarisation angle (PA) values with 1-$\sigma$ error bars referenced to infinite frequency and rotated by an arbitrary angle.
\textbf{b}, The total (I, black), linear (L, red) and circular (V, blue) intensities across the burst envelope. 
}
\label{fig:polarization}
\end{center}
\end{figure}

For FRB 20210206A, RM-synthesis results in an unambiguous RM detection. This detection is refined by applying a Stokes QU-fitting routine that directly fits for the modulation between Stokes $Q$ and $U$ from Faraday rotation as well as the modulation between $U$ and $V$ introduced by an instrumental delay between the two linear polarisations. Optimal values are determined numerically through Nested Sampling, a Monte Carlo method that seeks to optimize the likelihood function given a model and data. Further details on the CHIME/FRB polarisation analysis pipeline are presented elsewhere.\cite{Mckinven21}
The instrumental delay, once fitted for, can be used to produce a delay-corrected spectrum. The RMs produced from these two independent methods agree within the measurement uncertainties. An $\mathrm{RM} = +193.6(1)$\,rad\,m$^{-2}$ is given by QU-fitting and it is used to produce the polarized burst profile shown in Extended Data Figure~\ref{fig:polarization}.
We estimate a Galactic contribution of $\mathrm{RM_{MW}} = -150(33)$\,rad\,m$^{-2}$ along the sightline of FRB 20210206A,\cite{hab+2021} suggesting a significant extragalactic source of Faraday rotation. 

FRB 20210213A, conversely, displays an absence of polarised signal. Applying RM-synthesis produces no clear RM detection over the range $\mathrm{-2000 \lesssim RM \lesssim 2000 \; rad \, m^{-2}}$. 
For $\mathrm{|RM|}$ values beyond this range, bandwidth depolarisation from Faraday rotation within a single frequency channel becomes significant at the native channelization of CHIME/FRB baseband data.\cite{Mckinven21}
We developed an algorithm\cite{Mckinven21} that employs a phase-coherent method of correcting for bandwidth depolarisation in data for which the electric field phase is retained.\cite{vanStraten2002}
Using this method, we search out to $\mathrm{|RM|}$ values as large as $\mathrm{10^6 \; rad \, m^{-2}}$ by applying coherent de-rotation to a sparse grid of trial RMs followed by an incoherent search at neighbouring RM values. In principle, this method extends detectable RMs to arbitrarily large values. In practice, artifacts introduced in the channelization of CHIME/FRB baseband data reduce sensitivity to polarized signals at larger $\mathrm{|RM|}$ values. Given the low S/N of this burst, it remains possible that this event displays a large $\mathrm{|RM|}$ that simply goes undetected due to the deleterious effect of the channelization procedure. We test this possibility by using simulated data to determine the loss of S/N with increasing RM. Using a simulated burst with properties similar to that of FRB 20210213A (e.g., S/N, subband), we evaluate the performance of our coherent de-rotation algorithm over a range of RM values. We find no significant loss of polarised signal out to RM values as large as $\mathrm{|RM| \sim 200, 000 \; rad \, m^{-2}}$. Therefore, if this event does indeed display an RM within this range, a significant fraction of the signal must be unpolarised ($\gtrsim 50\%$) for us to not detect it given the S/N of the event. Conversely, we rule out the possibility of $\mathrm{|RM|}$ values larger than $\mathrm{\sim 200,000 \; rad \, m^{-2}}$ by detecting a lack of splitting in the burst morphology potentially caused by extreme RM values.\cite{sc19}

We note that ionospheric RM has not been corrected for in our analysis, but it does not exceed a few $\mathrm{rad \, m^{-2}}$.

\section*{Model of gravitational lensing}
As a possible explanation to the magnification that would be necessary to observe a radio pulsar located in another galaxy, we explore the observability of a pulsar that is gravitationally micro-lensed. 
In this model, pulses from a radio pulsar in a binary system are magnified in intensity by the gravitational field of its companion.
The magnification would be modulated in time such that the signal from the pulsar would be convolved with a bell-shaped curve. The pulse profiles detected for the three FRBs presented here are qualitatively consistent with this morphology.

We consider a binary system with a pulsar of mass $M_s = 1\,M_\odot$ and explore the parameter space of lensing masses, system alignment, and orbital separations to explain the properties of the three FRBs. 

We use a test pulsar at a distance of 1\,Gpc (approximately the DM-inferred distance of FRB 20210213A) emitting periodic pulses with a luminosity of 1\,Jy\,kpc$^2$, comparable to Galactic radio pulsars. 
Without any magnification, such a pulsar would be observed on Earth with a peak flux of $\sim 10^{-11}$\,Jy.
Therefore, a magnification $\mu \gtrsim 10^{11}$ is needed to explain the fluxes of $\sim 1$\,Jy observed for the three FRBs here reported.

We constrain the parameter space by first requiring the lens mass $M_\textrm{lens}$ to be able to magnify the pulsar\cite{1992grleS}
\begin{equation}\label{eq:lens_mass}
    \mu \lesssim  \frac{4 \pi G M_\textrm{lens}  f }{c^3},
\end{equation}
where $G$ is the gravitational constant and $f$ is the observed frequency.

\begin{figure}
\begin{center}
\includegraphics[width=.7\textwidth]{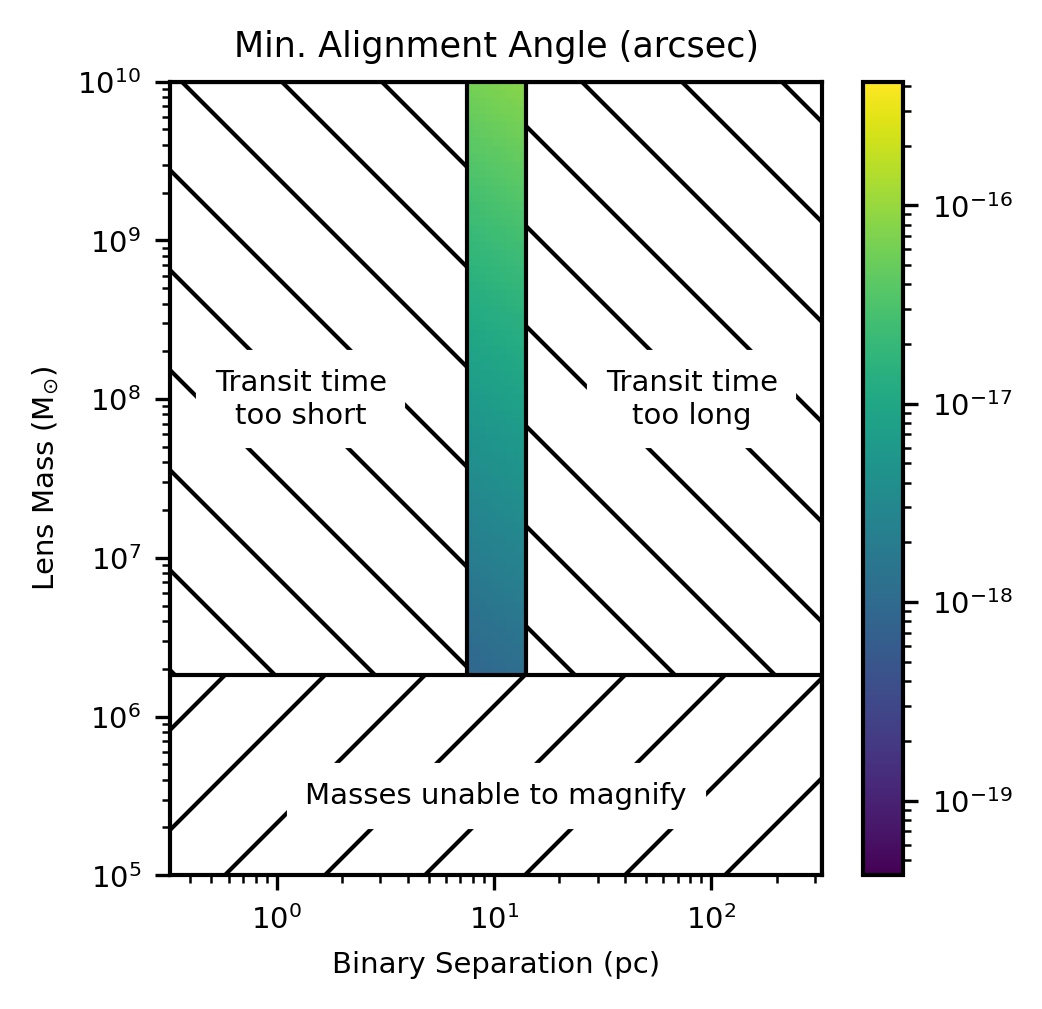}
\caption{
\textbf{Parameters of a binary system producing a radio signal compatible with the FRBs presented here through gravitational lensing.}
The system, located at 1\,Gpc, contains a 1\,$M_\odot$ pulsar emitting 1\,Jy pulses which are lensed by its companion.
The allowed parameter space is shown with brighter colors as a function of the minimum alignment angle (color-coded), companion mass and separation of the binary system.
}
\label{fig:lensing}
\end{center}
\end{figure}

The orbital separation of the binary system $D_{ls}$ and its alignment are constrained by requiring that we only observe the magnification curve for the duration of the event $\Delta t$.
The minimum alignment angle $\beta$ of the binary system can be calculated as\cite{1992grleS}
\begin{equation}\label{eq:alignment_angle1}
    \mu = \frac{\beta^2 + 2\theta_E^2}{\beta\sqrt{\beta^2 + 4\theta_E^2}},
\end{equation}
where the Einstein angle $\theta_E$ is defined as
\begin{equation}
    \theta_E = \sqrt{\frac{4 G M_{lens}}{c^2} \frac{D_{ls}}{D_{os} D_{ol}}}.
\end{equation}
$D_{os}$ is the distance between the observer and the source, $D_{ol}$ is the distance between the observer and the lens, and $D_{ls}$ is the orbital separation that we want to constrain.
$\beta$ is further constrained by
\begin{equation}\label{eq:alignment_angle2}
    \beta \gtrsim \omega \Delta t,
\end{equation}
where $\omega$ is the orbital angular velocity of the lens, defined as
\begin{equation}
    \omega  =\sqrt{\frac{G M_{lens}^3}{D_{ls}^3 (M_s + M_{lens})^2 }},
\end{equation}
and $\Delta t$ is the duration of time in which the pulsar is magnified enough to be detected above the noise floor.
We use Eqs.~\ref{eq:lens_mass}, \ref{eq:alignment_angle1}, and \ref{eq:alignment_angle2} to constrain the properties of the binary system and plot the parameter space shown in Extended Data Figure~\ref{fig:lensing}.

The orbital inclination of the lensing system would need to be aligned to the line of sight within $10^{-17} - 10^{-18}$ arcseconds, depending on the lens mass. 
The small required alignment angle implies a low probability to observe such an event despite the large trial factor provided by the number of galaxies within a Gpc distance. In addition, in this scenario, we would expect to detect a larger number of FRBs from gravitationally lensed pulsars in the nearby universe since they would require a lower magnification. Therefore, the absence of multi-peaked, periodic FRBs with a small DM excess\cite{1stcatalog} implies that it is unlikely that we have detected a gravitationally micro-lensed pulsar of average luminosity at a distance of 1 Gpc.

\section*{Model of merging compact objects}
A different model that could produce periodic FRBs considers merging neutron stars that emit the FRB signal.
One possible interaction of merging neutron stars to produce periodic FRBs is through a unipolar inductor process where the companion orbiting through the magnetic field acts as a conductor driving a current loop. The latter accelerates electrons and positrons to emit curvature radiation\cite{piro2012, mingarelli2015, wang2016, wang2018} in orbital frequencies ranging from few Hz to kHz, corresponding to orbital separations of 10--1000\,km in the binary neutron star case. Another proposed mechanism to extract energy is through the magnetic braking and spin-orbital synchronization of merging binary neutron stars.\cite{hansen2001, totani2013} The coherent emission is hypothesized to arise from the magnetosphere in a manner roughly similar to isolated pulsars as the rotation rate of one of the neutron stars rapidly increases or decreases to synchronize with the binary rotation. 
In such a case, FRB emission may show multiple peaks corresponding to a favourable orbital phase for a range of orbital periods.

The loss of angular momentum and energy through gravitational wave radiation causes the compact object binary orbits to decay with a predictable relation between the orbital angular frequency $\omega$ and time $t$. We consider the equation\cite{blanchet1995} for the instantaneous orbital angular frequency derivative (with $c$=$G$=1)
\begin{equation}\label{omega_dot}
\dot{\omega} = \frac{96}{5} \eta m^{5/3}\omega^{11/3} \left[ 1 - \left(\frac{743}{336} + \frac{11}{4}\eta\right)(m\omega)^{2/3}  \\
+ 4\pi m\omega + \left(\frac{34103}{18144} + \frac{13661}{2016}\eta + \frac{59}{18}\eta^2 \right)(m\omega)^{4/3}\right].        
\end{equation}
Here, $m = m_1 + m_2$ and $\eta = m_1m_2/m^2$ are the total mass and reduced mass, respectively, of two components of mass $m_1$ and $m_2$. We have assumed that spin-orbit and spin-spin coupling are negligible. 
Since the orbital angular frequency is related to the observed pulse period by
\begin{equation}
    P = \frac{2\pi}{\omega},
\end{equation}
it follows that 
\begin{equation}
    \dot{P} = -\frac{2\pi\dot{\omega}}{\omega^2}.
\end{equation}

For a given set of trial masses, $(m_1, m_2)$, we integrated numerically Eq.~(\ref{omega_dot}) twice to calculate the orbital phase $\phi$, i.e. we go from  $\dot{\omega}(\omega, t) \rightarrow \omega(t) \rightarrow \phi(t)$. The initial orbital period was chosen to be larger than the periods measured for these FRBs and the system was evolved until $\omega=2\pi\times10^6\,\mathrm{rad\,s^{-1}}$, very close to the final merger. We numerically inverted $\phi(t)$ to get $t(\phi)$, the time of passage of the components through a specific orbital phase. We use this to fit the ToAs through the same procedure used to study the significance of the periodicity. We modified Eq.~(\ref{eq:nt_regression}) as 
\begin{equation}
    t_i = t(2\pi n_i + \phi_0) + r_i,
\end{equation}
where $\phi_0$ is an arbitrary initial phase and the integer-valued $n_i$ vector is defined in Eq.~(\ref{eq:nj12}). We fixed one of the components to be a neutron star with $m_1 = 1.4\,\mathrm{M_\odot}$ and then searched the parameter space of ($m_2,\,\phi_0,\, n_i$) to minimize the root-mean-square of the residuals $r_i$. 
We find that FRB 20210206A cannot be explained by a merging neutron star model as the expected period derivative $\dot{P}\sim5\times10^{12}\,\mathrm{s\,s^{-1}}$ is not consistent with the observed peak separation. 
On the other hand, the ToAs for FRB 20191221A and FRB 20210213A are well-fit by this model for a broad range of mass $m_2$ ($0.1-6\,\mathrm{M_\odot}$ and $1.3-6\,\mathrm{M_\odot}$, respectively) and the same allowed space of `gap-vectors' $n_i$ as the previous fits. The putative separation of the neutron stars from their companions for these orbital fits would be $10^3$\,km and $10^2$\,km, respectively. These systems would be extremely short-lived. The timescale for these systems to merge, from the FRB event time is about $\sim10^2-10^4$\,s and 0.1--1\,s, respectively, depending on the companion masses. An eventual future detection of repeating bursts from these FRB sources would disprove this model.

\end{methods}

\clearpage
\printbibliography[segment=\therefsegment,heading=subbibliography,filter=notother]

\begin{addendum}
\item[Data availability]
The data used in this paper is stored into Hierarchical Data Format 5 files available at \url{https://doi.org/10.11570/22.0003}.

\item[Code availability]
The code used to model the signal from the sources presented in this publication, calculate their periodicities and plot the results is available at \url{https://doi.org/10.11570/22.0003}, together with the algorithms to calculate the $\hat S$ score and the Rayleigh statistic $Z_1^{2}$ used to estimate the significance of the periodicities.

\item
\newcommand{\genacks}{
We acknowledge that CHIME is located on the traditional, ancestral, and unceded territory of the Syilx/Okanagan people.
We thank the Dominion Radio Astrophysical Observatory, operated by the National Research Council Canada, for gracious hospitality and expertise. 
CHIME is funded by a grant from the Canada Foundation for Innovation (CFI) 2012 Leading Edge Fund (Project 31170) and by contributions from the provinces of British Columbia, Qu\'ebec and Ontario. The CHIME/FRB Project is funded by a grant from the CFI 2015 Innovation Fund (Project 33213) and by contributions from the provinces of British Columbia and Qu\'ebec, and by the Dunlap Institute for Astronomy and Astrophysics at the University of Toronto. Additional support was provided by the Canadian Institute for Advanced Research (CIFAR), McGill University and the McGill Space Institute via the Trottier Family Foundation, and the University of British Columbia.
The Dunlap Institute is funded through an endowment established by the David Dunlap family and the University of Toronto. 
Research at Perimeter Institute is supported by the Government of Canada through Industry Canada and by the Province of Ontario through the Ministry of Research \& Innovation. 
The National Radio Astronomy Observatory is a facility of the National Science Foundation (NSF) operated under cooperative agreement by Associated Universities, Inc. 
FRB research at UBC is supported by an NSERC Discovery Grant and by the Canadian Institute for Advanced Research. 
The CHIME/FRB baseband system is funded in part by a Canada Foundation for Innovation John R. Evans Leaders Fund award to I.H.S.
}
\genacks
\allacks
    \item[Author Contributions]
    All authors from CHIME/FRB collaboration played either leadership or significant supporting roles in one or more of: the management, development and construction of the CHIME telescope, the CHIME/FRB instrument and the CHIME/FRB software data pipeline, the commissioning and operations of the CHIME/FRB instrument, the data analysis and preparation of this manuscript.
 	All authors from CHIME collaboration played either leadership or significant supporting roles in the management, development and construction of the CHIME telescope.

    \textbf{Correspondence and requests for
        materials} should be addressed to Daniele Michilli\\(email: danielemichilli@gmail.com).

\end{addendum}


\end{document}